\documentclass[journal=jpcl,manuscript=extarticle,layout=twocolumn]{achemso}
\setkeys{acs}{maxauthors=0}

\let\titlefont\undefined
\makeatletter
\let\l@addto@macro\relax
\makeatother
\usepackage[fontsize=9pt]{scrextend}

\usepackage{graphicx}
\usepackage{bm}
\usepackage{xcolor}
\usepackage{siunitx}
\usepackage{physics}

\newcommand{\rd}{\mathrm{d}}

\usepackage{soul} 
\newcommand{\rev}[1]{\textcolor{black}{#1}} 
\newcommand{\onlinecite}[1]{\nocite{#1}\citenum{#1}}

\usepackage{amssymb}

\usepackage{xr-hyper}
\usepackage{hyperref}

\usepackage[symbol]{footmisc}

\usepackage{braket}
\usepackage{dcolumn} 
\newcolumntype{d}[1]{D{.}{.}{#1}} 

\title{Quantum Quality with Classical Cost: \textit{Ab Initio} Nonadiabatic Dynamics Simulations using the Mapping Approach to Surface Hopping}
\author{Jonathan R.\ Mannouch}
\affiliation{Hamburg Center for Ultrafast Imaging, Universit\"at Hamburg and the Max Planck Institute for the Structure and Dynamics of Matter, Luruper Chaussee 149, 22761 Hamburg, Germany}
\email{jonathan.mannouch@mpsd.mpg.de}
\author{Aaron Kelly}
\email{aaron.kelly@mpsd.mpg.de}
\affiliation{Hamburg Center for Ultrafast Imaging, Universit\"at Hamburg and the Max Planck Institute for the Structure and Dynamics of Matter, Luruper Chaussee 149, 22761 Hamburg, Germany}

\date{\today}

\makeatletter
\newcommand*{\addFileDependency}[1]{
  \typeout{(#1)}
  \@addtofilelist{#1}
  \IfFileExists{#1}{}{\typeout{No file #1.}}
}
\makeatother

\newcommand*{\myexternaldocument}[1]{
    \externaldocument{#1}
    \addFileDependency{#1.tex}
    \addFileDependency{#1.aux}
}

\myexternaldocument{si}

\begin{document}

\maketitle

\begin{abstract}
 Nonadiabatic dynamics methods are an essential tool for investigating photochemical processes. In the context of employing first principles electronic structure techniques, such simulations can be carried out in a practical manner using semiclassical trajectory-based methods or wave packet approaches. While all approaches applicable to first principles simulations are necessarily approximate, it is commonly thought that wave packet approaches offer inherent advantages over their semiclassical counterparts in terms of accuracy, and that this trait simply comes at a higher computational cost. Here we demonstrate that the mapping approach to surface hopping (MASH), a recently introduced trajectory-based nonadiabatic dynamics method, can be efficiently applied in tandem with \textit{ab initio} electronic structure. Our results even suggest that MASH may provide more accurate results than on--the--fly wave packet techniques, all at a much lower computational cost. 
\end{abstract}

\begin{center}
\includegraphics{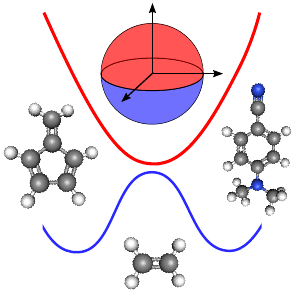}
\end{center}

Accurate computational simulations are crucial for understanding and interpreting experiments that investigate photoexcited molecular processes. Such systems are often high dimensional, involving multiple electronic potential energy surfaces and reaction channels, which makes constructing accurate parameterized models extremely challenging and time-consuming. Hence, performing \textit{ab initio} simulations `on--the--fly', such that the surfaces are computed in tandem with the propagation of the nuclear degrees of freedom, is often the only feasible option. In order to make such simulations practical, the number of electronic-structure calls per time step must be kept minimal, which requires dynamical approaches based on trajectories or localized nuclear basis functions.

Most dynamics approaches have not been specifically tested for \textit{\textit{ab initio}} simulations. Instead, these methods are commonly benchmarked on simplified models for which numerically exact quantum results can be generated.\cite{linearized,spinmap,multispin,spinPLDM1} However it is often unclear how far the conclusions from these simplified models can be extended to more realistic systems. Recently a test set of photoexcited molecular systems have been proposed as a benchmark for investigating the properties of different nonadiabatic dynamics algorithms using on--the--fly \textit{ab initio} electronic structure.\cite{Ibele2020} These three molecules -- ethylene, 4-(N,N-dimethylamino)benzonitrile (DMABN), and fulvene -- were initially chosen because their dynamics show similarities to the three so-called Tully models,\cite{Tully1990hopping} but their utility for benchmarking goes far beyond that. In particular, it is known that these systems give rise to different conical intersection topographies\cite{Ibele2020} and pathways for approaching the intersections,\cite{Gomez2024} therefore providing a rigorous test for any nonadiabatic dynamics method. This test set is currently gaining interest from the community, and it has already been the focus of a number of classical trajectory\cite{Weight2021,Ibele2020} and Gaussian wavepacket\cite{Ibele2020,Gomez2024} studies. In this work, we ascertain the accuracy and utility of a novel trajectory-based dynamics technique for performing \textit{ab initio} simulations, the mapping approach to surface hopping (MASH),\cite{Mannouch2023MASH} by applying it to simulate this molecular photochemistry test set and comparing our results with other more established methods. In order to determine a hierarchy in the accuracy of these approaches, we also compare with numerically exact quantum dynamics applied to linear vibronic coupling (LVC) models previously parameterized for these molecules.\cite{Gomez2024}

Before discussing the relaxation dynamics of the three photoexcited molecular systems, we give a brief overview of the dynamics approaches used in this work, with a particular emphasis on their similarities and differences. More information can be found in Refs.~\onlinecite{Subotnik2016review,Mannouch2023MASH,Ben-Nun2000,Curchod2018review}.

Fewest-switches surface hopping (FSSH)\cite{Tully1990hopping,Subotnik2016review} is the most popular independent-trajectory approach for simulating nonadiabatic dynamics in molecules. In FSSH, the nuclei are propagated according to (classical) Born-Oppenheimer molecular dynamics on a single surface 
and nonadiabatic transitions are described by stochastic changes in the `active' surface, called `hops'. The hopping probability is related to the rate of change of the underlying electronic wavefunction, which itself is propagated according to the associated time-dependent Schr\"odinger equation. One issue with altering the active surface in this way is that it can become inconsistent with the electronic wavefunction, leading to the so-called overcoherence error that is known to significantly degrade the accuracy of the obtained results.  To fix this, a number of decoherence corrections have been proposed, \cite{Hammes-Schiffer1994,Bittner1995,Jasper2005,Granucci2010,Subotnik2011AFSSH,Shenvi2011,Subotnik2011,Jaeger2012,Vindel2021} which sporadically reset the wavefunction to the current active surface. While not guaranteed,\cite{Schwerdtfeger2014ET} it is generally accepted that decoherence corrections lead to an improvement in the accuracy of the obtained results in the majority of cases.

Despite the substantial progress that has been made in understanding many foundational aspects of FSSH,\cite{Kapral2016,Subotnik2013} a number of variants of the FSSH dynamics algorithm are nevertheless still widely used in the literature. The main aspect that differs between most FSSH algorithms lies in the way that the nuclear velocities are rescaled at a hop. While it is generally agreed upon that rescaling along the nonadiabatic coupling vector (NACV) is the correct thing to do,\cite{Pechukas1969scattering2,Herman1984MomentumReversal,Hammes-Schiffer1994,Toldo2024} many other schemes are used in practical implementations of FSSH.\cite{Toldo2024} In particular, rescaling all degrees of freedom equally, which is often referred to as rescaling `along the velocity vector', is probably the most commonly used.\cite{Plasser2014,Suchan2020,Vacher2024azomethane} Additionally, an upward hop must be aborted if there is insufficient nuclear kinetic energy, often referred to as a `frustrated hop'. The nuclear velocity along the NACV is reflected at a frustrated hop in many FSSH implementations,\cite{Hammes-Schiffer1994} although other suggestions have been made,\cite{Muller1997,Jasper2003,Jain2015hopping2} along with those that try to avoid frustrated hops altogether.\cite{Fang1999,Jasper2002}

The mapping approach to surface hopping (MASH)\cite{Mannouch2023MASH} is a recently proposed independent-trajectory approach that alleviates the problems of FSSH by utilizing the best features of mapping-based semiclassical trajectories\cite{Meyer1979nonadiabatic,Stock1997mapping,spinmap,multispin} in a surface hopping algorithm. In many aspects the algorithm is identical to FSSH, but it contains the following key differences. In MASH, the active surface is not an additional parameter within the theory, but is uniquely determined from the electronic wavefunction. For two-state problems, this corresponds to selecting the surface for which the electronic wavefunction has the largest associated probability. In addition, the stochastic nature of hops in FSSH is replaced by a fully deterministic dynamics which guarantees that the electronic wavefunction and the active surface remain consistent. The overcoherence problem is therefore resolved in MASH without the need for ad hoc decoherence corrections. For example, MASH accurately captures nonadiabatic thermal rates,\cite{MASHrates} whereas decoherence corrections are known to be needed for the analogous FSSH simulations.\cite{Landry2011hopping,Landry2012hopping,Jain2015hopping1,Jain2015hopping2,Falk2014FSSHFriction} 

In fact, MASH gives a unique prescription for all aspects of the simulation algorithm, including the velocity rescaling and treatment of frustrated hops. In agreement with what many have suggested for FSSH,\cite{Pechukas1969scattering2,Herman1984MomentumReversal,Hammes-Schiffer1994,Toldo2024} MASH determines that the velocity must be rescaled along the NACV at a hop and reflected in the case of a frustrated hop, in order for the approach to reproduce the short-time behaviour of exact quantum dynamics. In particular, the exact quantum-mechanical equation of motion for the so-called `kinematic momentum'\cite{Cotton2017mapping} for mode $j$ depends on the Born-Oppenheimer and nonadiabatic contributions to the nuclear force, which are given by the two terms on the right-hand side of the following equation
\begin{equation}
\label{eq:force}
\frac{\rd}{\rd t}\braket{\hat{p}_{j}}=-\sum_{\lambda}\Braket{\frac{\partial V_{\lambda}}{\partial q_{j}}\hat{P}_{\lambda}}+2\sum_{\lambda}\sum_{\mu>\lambda}\Braket{(V_{\mu}-V_{\lambda})d^{(\mu,\lambda)}_{j}\hat{\sigma}^{(\mu,\lambda)}_{x}} ,
\end{equation}
where $V_{\lambda}$ is the Born-Oppenheimer surface for state $\ket{\psi_{\lambda}}$, $\hat{P}_{\lambda}=\ket{\psi_{\lambda}}\bra{\psi_{\lambda}}$ is the associated electronic population operator, $d_{j}^{(\mu,\lambda)}$ is the NACV between states $\ket{\psi_{\mu}}$ and $\ket{\psi_{\lambda}}$ and $\hat{\sigma}_{x}^{(\mu,\lambda)}=\ket{\psi_{\mu}}\bra{\psi_{\lambda}}+\ket{\psi_{\lambda}}\bra{\psi_{\mu}}$ is the associated electronic coherence operator. More details associated with this formula are given in the supporting information (SI). MASH is constructed to describe the Born-Oppenheimer (adiabatic) force through the active surface and the nonadiabatic force through the velocity rescaling performed along the NACV. In addition, MASH has already been benchmarked on a range of model systems,\cite{MASH,MASH_thermalization,MASHrates} where it was regularly found to offer improvements over FSSH. 

To summarize the above discussion, a hierarchy of the surface hopping approaches can be established according to their expected accuracy. Firstly, it is expected that surface hopping approaches that perform the velocity rescaling along the velocity vector (FSSH-vel) will be less accurate than those that perform the velocity rescaling along the NACV (FSSH-nacv), due to the fact that the later can describe dynamical effects arising from the nonadiabatic force. Secondly, fewest-switches surface hopping approaches that incorporate a decoherence correction (dFSSH) are expected to be more accurate than those that do not (FSSH), because it is important to impose consistency between the active surface and the electronic wavefunction. Finally, MASH is expected to be either just as accurate or more accurate than dFSSH-nacv. This hierarchy will be helpful for determining what is likely to be the correct dynamics in \textit{ab initio} photochemical simulations, where numerically exact quantum dynamics is not obtainable.

A completely different approach for describing nonadiabatic transitions compared to the quantum-classical approaches described above is taken in \textit{ab initio} multiple spawning (AIMS)\cite{Ben-Nun2000,Curchod2018review}. Based on a series of approximations to full multiple spawning,\cite{Ben-Nun1998,Mignolet2018} AIMS is a Gaussian wavepacket algorithm that was developed for performing on--the--fly \textit{ab initio} simulations. Gaussian basis functions are propagated classically on single Born-Oppenheimer surfaces and new Gaussians are `spawned' whenever the system enters a region of strong nonadiabatic coupling. The evolution of the Gaussian weights is then obtained by solving an approximate time-dependent Schr\"odinger equation spanned by the Gaussian basis. The main advantage of AIMS over the surface hopping approaches is its coupled trajectory nature, which goes beyond the independent trajectory approximation, albeit at an increased computational cost. On the other hand, given that AIMS only uses a minimal number of Gaussian basis functions, its accuracy will largely be determined by how effectively the Born-Oppenheimer forces are able to move the Gaussians into the correct regions of nuclear phase space. As a result, one way of improving upon AIMS is to instead propagate the Gaussians using equations of motion that ensure Eq.~(\ref{eq:force}) is satisfied, as is done for example in the variational multi-configuration Gaussian (vMCG) approach\cite{Richings2015vMCG}. Given that AIMS is not a benchmark method,\cite{AIMS_CI} the relative accuracy of AIMS compared to the quantum-classical approaches therefore depends on the relative severity of the independent trajectory approximation for realistic molecular simulations compared to how effectively the minimal set of Gaussian basis functions generated by AIMS spans the full support of the time-dependent wavefunction.

\begin{figure*}[t]
    \centering  \includegraphics[width=1.0\linewidth]{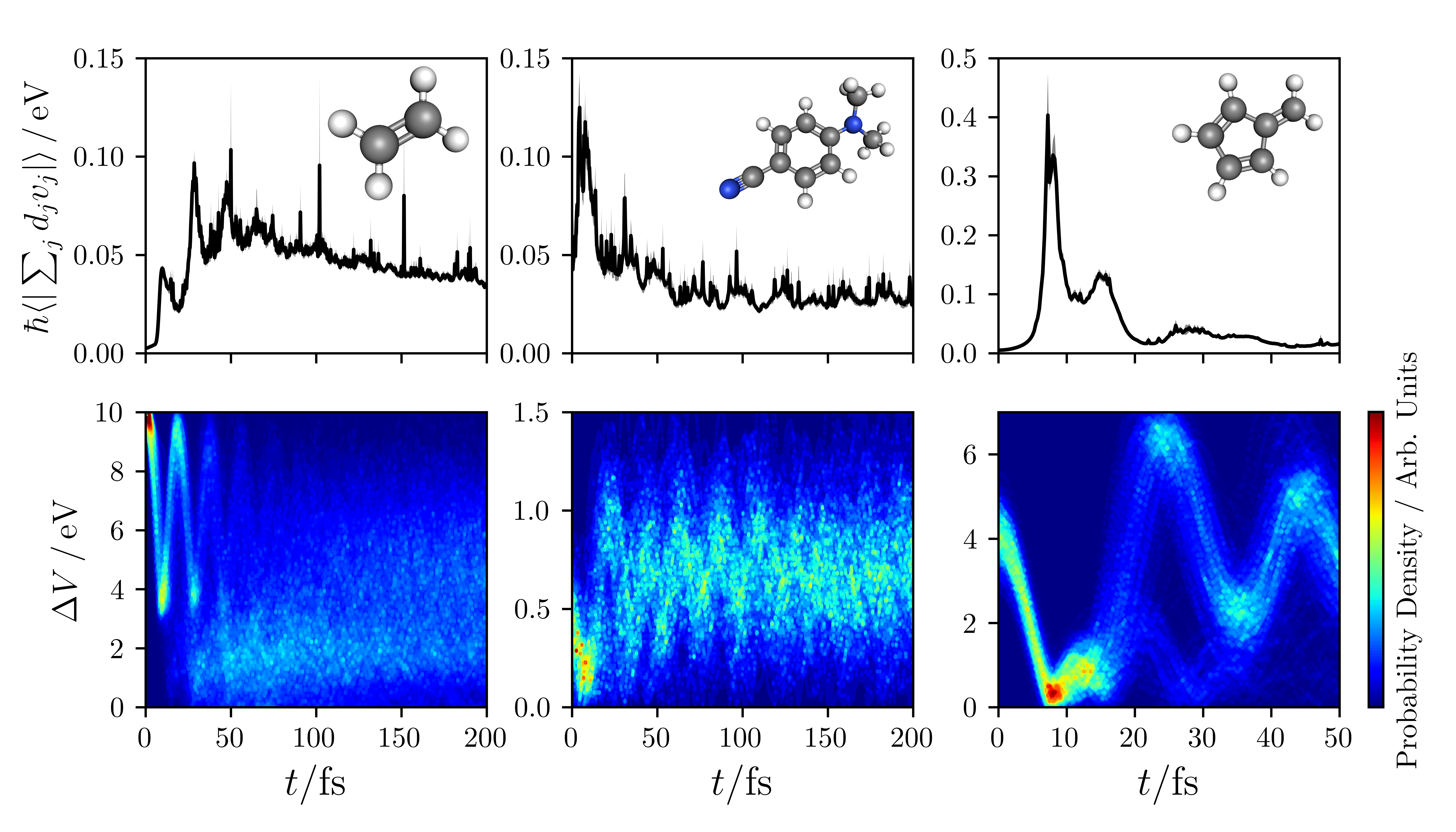}
    \caption{The magnitude of the effective nonadiabatic coupling, $|\sum_{j}d_{j}v_{j}|$, and the probability distribution of the time-dependent energy gap, $\Delta V=V_{+}-V_{-}$, between the two adiabatic states. These quantities are calculated for each system by averaging over the MASH trajectories.}
    \label{fig:model}
\end{figure*}

We now consider in more detail the relaxation dynamics of ethylene, DMABN, and fulvene, using the same electronic structure theory for each system as defined in Ref.~\onlinecite{Ibele2020}. The initial conditions are taken to be of the Franck-Condon type, with the electronic system on the upper of the two considered adiabats and the nuclear system in the ground vibrational state associated with the harmonic approximation to the electronic ground state potential. MOLPRO 2012\cite{Molpro2012} and GAUSSIAN 16\cite{g16} were used for the SA-CASSCF and LR-TDDFT electronic structure calculations, while the surface hopping and AIMS dynamics were performed using the SHARC 2.0\cite{SHARC2.0,Plasser2016,Mai2018SHARC} and AIMS/MOLPRO\cite{AIMS/MOLPRO} codes.  Files containing the electronic structure inputs and the ground-state geometries and frequencies for each system are provided in the supplementary material. Each system displays a different type of relaxation pathway involving conical intersections. 
One way that this excited-state relaxation dynamics can be visualized is through the probability density of the dynamical energy gap between the two Born-Oppenheimer surfaces, as presented in the second row of Fig.~\ref{fig:model} for the MASH trajectories. In these figures, the Franck-Condon region corresponds to a large value of $\Delta V$, while the conical intersection seams are about $\Delta V\approx 0$. Also given in the first row of the same figure is the effective nonadiabatic coupling between the states (i.e., the magnitude of the dot product of the NACV with the nuclear velocity) averaged over the same trajectories. 

Ethylene is known to have `indirect' access to its conical intersections,\cite{Gomez2024} meaning that the Franck-Condon region for the $\mathrm{S}_{0}\rightarrow\mathrm{S}_{1}$ transition lies far away from the conical intersection seams and the initial nuclear dynamics upon photoexcitation does not directly access them. A redistribution of the vibrational energy to the appropriate modes during a few vibrational time periods is necessary before the crossing region can be accessed, giving a delayed onset of the first nonadiabatic transitions to about $\approx 25$ fs. At later times, the majority of trajectories move away from the intersection region once they have reached the ground state. However the relatively slow decay in the coupling suggests that this process is relatively inefficient and that many trajectories may undergo multiple nonadiabatic transitions before remaining on the ground-state surface.

In contrast, upon a Franck-Condon type photoexcitation from the ground state to $\mathrm{S}_{2}$, DMABN has a very small energy gap between $\mathrm{S}_{2}$ and $\mathrm{S}_{1}$. As a result, the nonequilibrium dynamics in DMBAN were previously coined `immediate'\cite{Gomez2024} because the initial nuclear wavepacket is essentially on top of the conical intersection. The small initial energy gap between $\mathrm{S}_{2}$ and $\mathrm{S}_{1}$ also means that the dynamics remain close to the conical intersection seam for relatively long times, leading to the possibility that a large number of nonadiabatic transitions take place. This is consistent with the observation that the effective nonadiabatic coupling rapidly plateaus to a non-zero value. Experimentally it is known that relaxation to the $\mathrm{S}_{0}$ state occurs at much longer times predominantly through fluorescence, and so we exclude the possibility of nonadiabatic transitions between $\mathrm{S}_{1}$ and $\mathrm{S}_{0}$ in this study.

Despite the conical intersection seams in fulvene being further away from the Frank-Condon region than in DMABN, the initial motion of the nuclear wavepacket does still allow direct access to the crossing region,\cite{Gomez2024} in contrast to ethylene. The distinct feature of fulvene is that trajectories first pass through a sloped conical intersection seam driven by a stretch in the $\mathrm{C}\!\!\!=\!\!\!\mathrm{CH}_{2}$ moiety\cite{Ibele2020}, and then part of the wavepacket is reflected back through the crossing region at $\approx 15$ fs. This system therefore provides a useful test for how well different nonadiabatic dynamics approaches can correctly describe recrossing phenomena. While a peaked conical intersection seam also exists in fulvene, driven by a twist in the  $\mathrm{C}\!\!=\!\!\mathrm{CH}_{2}$ moiety, we find that this intersection is not accessed until much later times. 

\begin{figure*}[t]
    \centering
    \includegraphics[width=1.0\linewidth]{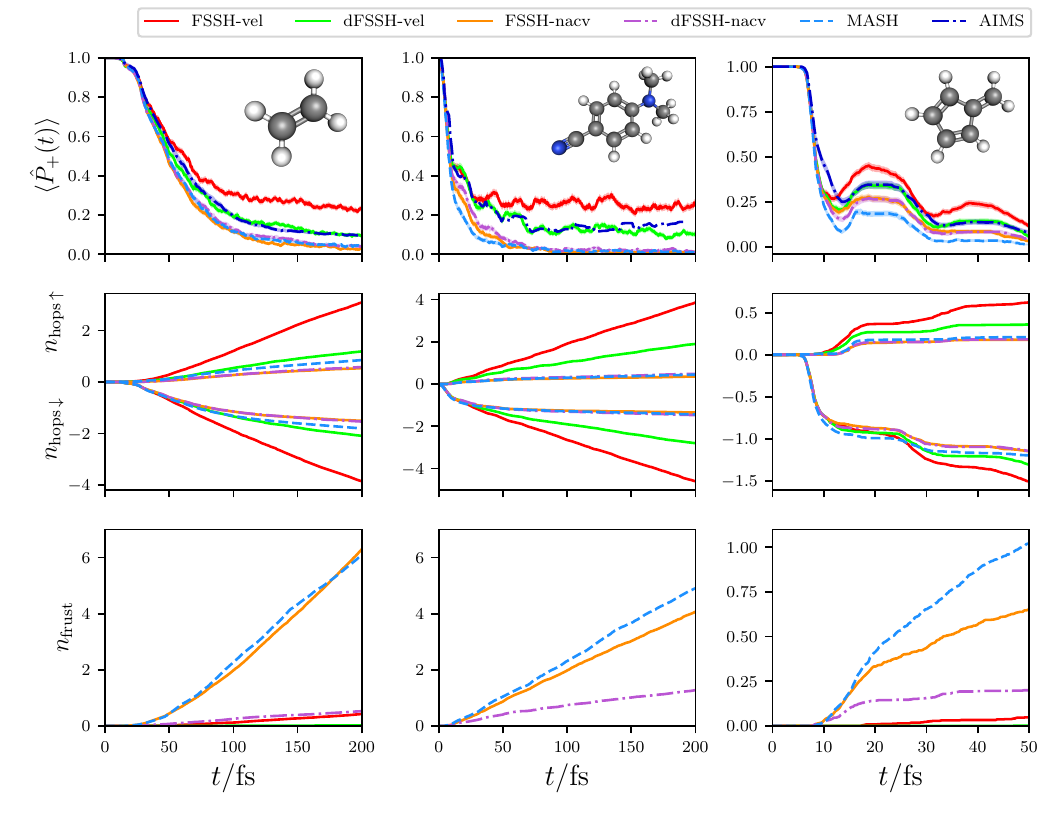}
    \caption{The electronic population of the upper adiabatic state ($\braket{\hat{P}_{+}(t)}$) for ethylene, DMABN and fulvene, computed for a wide variety of different methods. The width of the shading represents twice the standard error, which is less than the line width if not visible. The AIMS result for DMABN was obtained from Ref.~\protect\onlinecite{Curchod2017}. For the surface hopping approaches, the average number of upward hops ($n_{\mathrm{hops}\,\uparrow}$) downward hops ($n_{\mathrm{hops}\,\downarrow}$) and frustrated hops ($n_{\mathrm{frust}}$) are also given.}
    \label{fig:populations}
\end{figure*}

We next analyze the time evolution of the electronic excited state populations, shown in the first row of Fig.~\ref{fig:populations}. In order to help analyze the differences in the obtained populations from different surface hopping algorithms, the number of allowed and frustrated hops are also given in the same figure. In particular, the adiabatic populations can be exactly reproduced from the difference in the number of downward and upward hops. Despite the differing properties of the conical intersections involved in these three systems, the general trend in the results for each method is largely the same. 

First, we observe that the direction in which the velocity rescaling is performed in surface hopping approaches results in a significant quantitative difference in the obtained electronic populations. For FSSH performed by rescaling along the velocity vector, dFSSH-vel deviates significantly from FSSH-vel, suggesting that the decoherence error in this case is large. In contrast, for the surface hopping results where the velocity rescaling is applied along the NACV, FSSH-nacv and dFSSH-nacv are almost identical in all cases. This can be understood from the fact that in all systems, FSSH-vel trajectories undergo a larger number of allowed hops than FSSH-nacv, making it more likely that the electronic wavefunction can become inconsistent with the active surface in the former. Decoherence corrections are therefore more necessary in FSSH-vel for reimposing consistency. 

These differences between the algorithms can be further explained by considering the amount of nuclear kinetic energy available for promoting upward hops. In the case of rescaling along the velocity vector, the nuclear kinetic energy of the entire molecule is available for inducing electronic transitions, making almost all attempted hops energetically allowed. However, this behaviour is unphysical because not all of the nuclear degrees of freedom are directly coupled to the electronic transition.\cite{Carof2017,Braun2022} In contrast, the NACV rescaling direction ensures that only the kinetic energy associated with directly coupled modes is considered; this is significantly less than the nuclear kinetic energy of the entire molecule, making upward hops more likely to be energetically forbidden and therefore frustrated. The associated electronic population for the upper adiabat is therefore significantly lower for dFSSH-nacv than for dFSSH-vel. This effect has been observed in other theoretical studies,\cite{Plasser2019} including those on ethylene\cite{Barbatti2021} and fulvene.\cite{Ibele2020,Toldo2024}

In particular, this leads to qualitatively different dynamics than was previously predicted for DMABN. One of the main reasons that DMABN was previously suggested as a good candidate benchmark system was that its dynamics were expected to produce similar features to Tully's model II.\cite{Ibele2020} It was known that the adiabatic potential energy surfaces remain close in energy throughout the dynamics (as also illustrated in Fig.~\ref{fig:model}), suggesting that repeated electronic transitions between the surfaces would occur. While this is indeed observed when rescaling along the velocity vector, the observed dynamics when rescaling along the NACV instead involves a single rapid transition to the lower adiabatic state, where the system remains indefinitely. While the potential energy surfaces in DMABN do remain close together in energy relative to the kinetic energy of the entire molecule, they do not relative to the kinetic energy along the NACV. 

Of the three systems, fulvene is particularly interesting because the MASH result significantly deviates from both FSSH-nacv and dFSSH-nacv. There is also a noticeable difference between these results for DMABN too, although the difference is much smaller. Figure~\ref{fig:populations} shows that this deviation between the MASH and dFSSH-nacv electronic populations is predominately due to the larger number of frustrated hops in MASH, which leads to the MASH electronic populations being slightly lower than those of dFSSH-nacv. 

What is also interesting about Fig.~\ref{fig:populations} is that the electronic populations obtained by AIMS are seen to be almost identical to those obtained with dFSSH-vel. Firstly, this suggests that the independent trajectory approximation that underpins all surface hopping approaches is valid for these systems. While there are small deviations between the dFSSH-vel and AIMS results around 10 fs in fulvene and towards longer times in DMABN, we note that these differences are relatively minor compared to the more significant discrepancies observed between other algorithms. More importantly, the fact that AIMS and dFSSH-vel agree so well suggests that AIMS may not be describing the effect of the nonadiabatic force, which is an effect that is also neglected in dFSSH-vel. This finding is not so surprising in the case of DMABN, where the AIMS simulation did rescale the average velocity of spawned Gaussians along the velocity vector to ensure classical energy conservation.\cite{Curchod2017} However in the AIMS simulations for ethylene and fulvene, this rescaling was performed along the NACV, which at least for surface hopping algorithms is sufficient to correctly describe the nonadiabatic force. Our findings are however consistent with the observation that, unlike for surface hopping approaches, the velocity rescaling direction in AIMS is found to make almost no difference to the obtained results.\cite{Ibele_Thesis} 
While this point certainly requires further investigation, it does however suggest that the AIMS population dynamics may be less accurate than the best surface hopping algorithms in these cases.   

In previous work, other algorithms have also been tested on these systems using the same initial conditions and electronic structure methods. In Fig.~S4 in the SI, we compare AIMS and MASH to various flavours of the symmetrical quasi-classical (SQC)\cite{Miller2016Faraday} method for the population dynamics of ethylene. The SQC results were obtained from Ref.~\onlinecite{Weight2021}. With the exception of non-gamma corrected SQC using square windows, all of the other SQC approaches give results somewhere in between AIMS and MASH. Given that the SQC results contain large statistical error, it is however hard to precisely ascertain the relative accuracies of the various approaches. While the surface hopping algorithms were performed with slightly more trajectories (1000) than the SQC approaches (500), given the observed noise in the data, we estimate that at least an order or magnitude more SQC trajectories would be needed to obtain a similar level of convergence as the surface hopping results. This highlights one of the main advantages of the surface hopping approaches (including MASH) over mean-field mapping-based approaches, in that they require significantly fewer trajectories to converge the results. 

\begin{figure*}[t]
    \centering
    \includegraphics[width=1.0\linewidth]{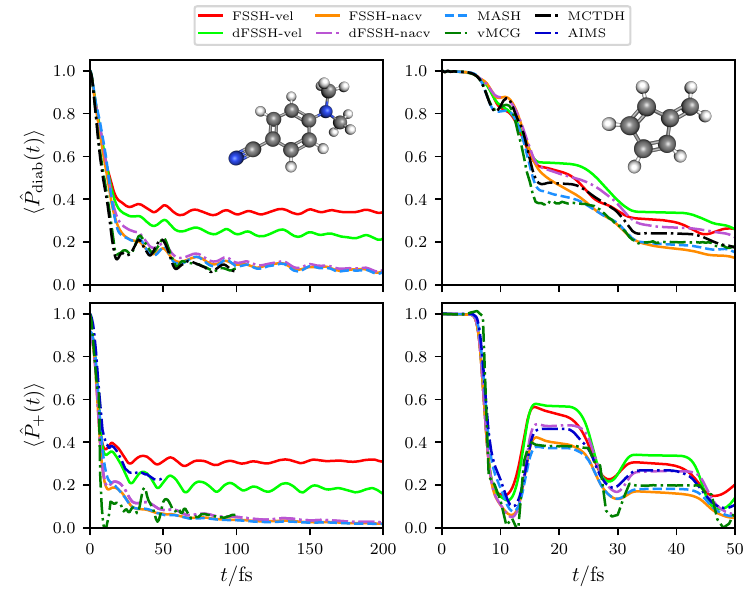}
    \caption{The electronic populations of the upper adiabatic state ($\braket{\hat{P}_{+}(t)}$) and the diabatic state that coincides with this adiabat at the Franck-Condon geometry ($\braket{\hat{P}_{\mathrm{diab}}(t)}$), obtained for the linear vibronic coupling (LVC) models that were constructed for DMABN and fulvene in Ref.~\protect\onlinecite{Gomez2024}. The MCTDH and vMCG results were also obtained from Ref.~\protect\onlinecite{Gomez2024}. We do not consider the analogous LVC model for ethylene, as it was found to not give rise to any electronic population transfer.}
    \label{fig:populations_lvc}
\end{figure*}

To complement the above comparison of the various dynamics approaches in the \textit{ab initio} case, we also perform simulations for linear vibronic coupling (LVC) models fit to the electronic structure data for DMABN and fulvene. These LVC models have already been used to perform numerically exact quantum dynamics using the multi-configuration time-dependent Hartree (MCTDH) approach, along with some vMCG calculations.\cite{Gomez2024} Given that it is significantly easier to compute diabatic populations with MCTDH, and adiabatic populations with AIMS, we provide both quantities in Fig.~\ref{fig:populations_lvc}. More details regarding the LVC model calculations can be found in the SI. For the diabatic populations in the DMABN model, vMCG and MCTDH are essentially indistinguishable from one another and MASH, FSSH-nacv and dFSSH-nacv produce significantly more accurate results than FSSH-vel and dFSSH-vel. While the situation is less clear cut in the fulvene model, both MASH and vMCG very accurately match the MCTDH result up to $\approx 10$ fs and remain relatively close to it after that. For the adiabatic populations, the trend in the results for all of the approaches is largely the same as in the \textit{ab initio} simulations. \footnote{Interestingly, the fulvene model unlike in the \textit{ab initio} simulations seems to also provide a case for which applying decoherence corrections appears to make the FSSH results worse.} Most importantly, the vMCG adiabatic populations agree best with those from MASH and FSSH-nacv, further suggesting that these surface hopping approaches are performing the best among all of the `on--the--fly' approaches that we have tested. Given that the main difference between the vMCG and AIMS algorithms is how the Gaussians are propagated further suggests that this is the source of the error observed in the AIMS results.  

We now focus on comparing the best surface hopping approaches of FSSH-nacv, dFSSH-nacv and MASH with AIMS for nuclear observables, and we provide the FSSH-vel and dFSSH-vel results in the SI for completeness. Not all nuclear observables are particularly sensitive to the nonadiabaticity of the problem, however. For example one particularly interesting nuclear observable in the case of DMABN is the twist angle of the dimethylamino group. This is because the ground state structure is untwisted, while both of the minimum energy configurations of the $\mathrm{S}_{1}$ surface are twisted. In addition, there are two minimum energy conical intersections (MECIs) between $\mathrm{S}_{1}$ and $\mathrm{S}_{2}$, one of which is twisted and the other that is not.\cite{Curchod2017} It is therefore interesting to ask whether the onset of twisting in the dynamics is directly connected to the nonadiabatic transition. In Fig.~S3 in the SI, we give the dynamical twist angle computed using a selection of methods. The fact that the observed twist angles are within the statistical error for all methods suggests that the nonadiabatic transitions are occurring through the untwisted MECI and the observed twisting arises from the topology of the $\mathrm{S}_{1}$ surface. This is also in agreement with the conclusions of previous wavepacket and surface hopping simulations on DMABN.\cite{Du2015,Kochman2015,Curchod2017,Gomez2021}

\begin{figure*}[t]
    \centering
    \includegraphics[width=1.0\linewidth]{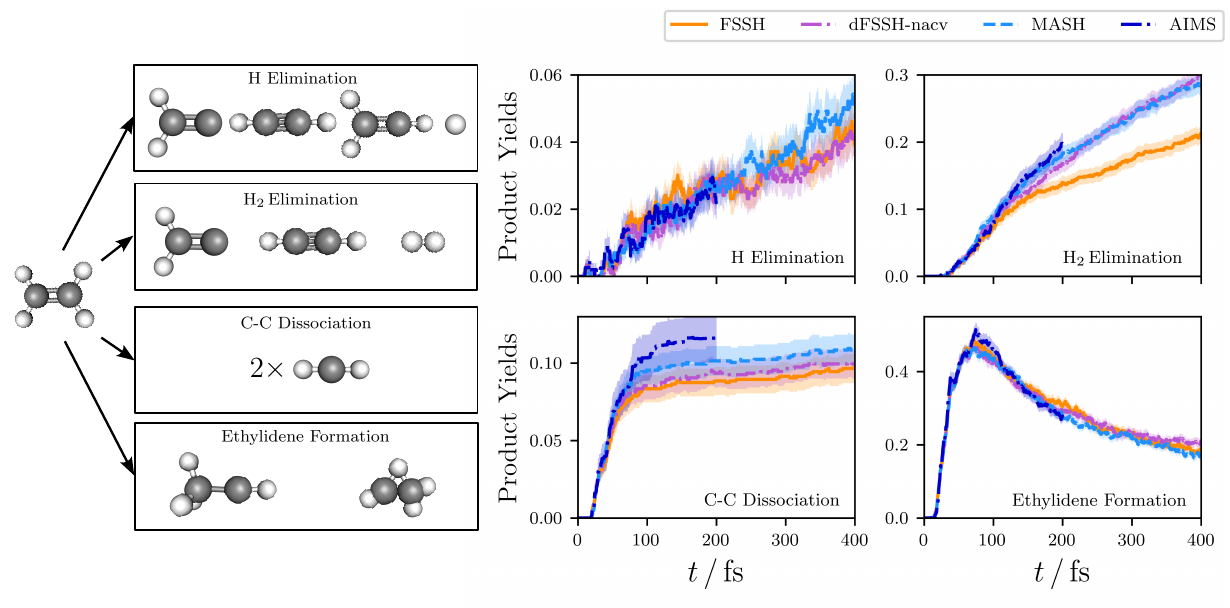}
    \caption{Dynamical product yields for the major products in the photodissociation of ethylene, calculated using various dynamical methods. The width of the shading represents twice the standard error.}
    \label{fig:yield}
\end{figure*}

The product yields in the photodissociation of ethylene are an example of nuclear observables that end up being more sensitive to the nonadiabaticity of the problem. In the following, we group the possible products according to the four different channels depicted in Fig.~\ref{fig:yield}.\cite{Barbatti2005,Miyazaki2023} All products observed in our computational simulations match those found in the corresponding experiments,\cite{Balko1992,Lin2000,Lee2004,Kosma2008,Allison2012} with the exception of the C-C dissociation process. As commented on in previous theoretical studies of ethylene,\cite{Miyazaki2023} the C-C dissociation is a result of the inadequacy of the basis set used in the electronic structure calculations, which gives a $\mathrm{S}_{0}-\mathrm{S}_{1}$ excitation energy of 10.2 eV at the Franck-Condon geometry, which is much larger than the experimental value of 7.6 eV,\cite{RobinBook} and most importantly significantly above the C-C bond energy of 7.7 eV.

The dynamical product yields calculated for a range of dynamical methods are given in Fig.~\ref{fig:yield}. Of the products considered, the yields of H elimination, C-C dissociation and ethylidene formation are largely the same for all the methods considered, at least within the statistical error. The geometries corresponding to ethylidene are known to match those of various conical intersection seams in the system.\cite{Levine2007,Miyazaki2023} The product yield for this process therefore qualitatively describes the initial approach and subsequent exit of the conical intersection regions during the dynamics, explaining why it also qualitatively matches the average effective coupling between the Born-Oppenheimer surfaces, as given in Fig.~\ref{fig:model}.

Of particular interest is the product yield for $\mathrm{H}_{2}$ elimination, where dFSSH-nacv, MASH and AIMS all give rise to an enhanced product yield over FSSH-nacv. Such behaviour is reminiscent of the use of surface hopping to calculate nonadiabatic thermal rates. In this case, the inconsistency error of FSSH is known to suppress the observed reaction rate and mean that the correct $\Delta^{2}$ scaling behaviour with respect to the diabatic coupling strength is not correctly reproduced.\cite{Landry2011hopping,Landry2012hopping,Jain2015hopping1,Jain2015hopping2,Falk2014FSSHFriction} More specifically in the thermal rate problem, it was found that at short times the two-hop trajectories between the ground-state reactant and product geometries were the ones that were responsible for the incorrectly suppressed reaction rate in FSSH.\cite{MASHrates} In the $\mathrm{H}_{2}$ elimination process, the trajectories associated with an odd number of hops are important, because the reaction proceeds from the excited to the ground electronic state. As a result, we find that it is the three- and five-hop trajectories\footnote{In analogy to the thermal rate problem,\cite{MASHrates} one may expect that only the one and three-hop trajectories are important for determining the rate. The reason that the 5-hop trajectories must also be considered is because the transient behaviour is now longer as a result of this being a nonequilibrium rate process.} that are the problem for FSSH. The advantage of AIMS and MASH is that the correct nonadiabatic rate is reproduced without the need for ad hoc decoherence corrections. Note that while performing AIMS simulations for long enough times to observe the reaction is expensive,\footnote{Because of this issue, we only run the AIMS simulations up to 200 fs. Some algorithmic advancements in AIMS have been designed to alleviate this issue,\cite{Lassmann2021} but we do not directly consider them here.} this is not the case for the independent MASH trajectories. 

In conclusion, we have applied a recently proposed independent-trajectory surface hopping approach, MASH, to perform \textit{ab initio} nonadiabatic dynamics simulations in molecules. We compared MASH with a set of well established methods over a series of two-state benchmark systems -- ethylene, DMABN and fulvene -- with the surfaces and couplings computed on-the-fly using various \textit{ab initio} electronic structure methods. Both electronic and nuclear observables were considered. 

Overall, MASH is likely to be the most suitable approach for performing \textit{ab initio} simulations in molecules due to its accuracy and efficiency. For electronic population observables, MASH is able to correctly describe the effects arising from the nonadiabatic force, which was seen to be absent in AIMS and the most commonly used velocity rescaling scheme of FSSH. Such findings were also corroborated by the use of LVC models that were parametrized by fitting to electronic structure data for these molecules, where comparison to numerically exact quantum dynamics was possible. Photodissociation product yields can also be accurately and robustly calculated with MASH, because it solves the inconsistency problem of FSSH without the need for ad hoc decoherence corrections.

Our analysis has uncovered a potential shortcoming within AIMS and it will be interesting to ascertain whether the nonadiabatic force can be correctly incorporated into the motion of the Gaussian basis functions within the AIMS algorithm. For the systems that we have tested here, the independent trajectory approximation is seen to be valid, so that even if AIMS can be fixed, the computational simplicity of the independent-trajectory nature of MASH is still likely to be superior. An interesting question is therefore whether photochemical (or other molecular) processes can be found where the independent trajectory approximation does break down and where coupled-trajectory approaches like AIMS do offer a distinct advantage.

There are some approximations that are shared by all of the `on--the--fly' approaches tested in this work. For example, all of them sample the initial nuclear phase-space variables from a Wigner distribution and propagate them classically. Therefore it is hard to ascertain how severe the potential issues of zero-point energy leakage and the lack of nuclear tunnelling are for these systems. MASH is guaranteed to thermalize correctly in the long-time limit with a classical nuclear bath,\cite{thermalization} but this is not guaranteed if some of the nuclei have a large zero-point energy. A direct dynamics version of vMCG has been used to study these systems, and it would be good to estimate the robustness of the classical nuclear approximation in these systems by comparing our results to this in future work. 

One problem of comparing different methods in \textit{ab initio} simulations is the difficulty in making sure that the results obtained with different code bases are compatible with one another. To this end, we have provided an extensive SI which addresses in detail the potential sources of discrepancies between different codes, as well as showing how the dynamics in SHARC\cite{SHARC2.0} and the AIMS/MOLPRO package\cite{AIMS/MOLPRO} can be made consistent with one another. For future benchmarking exercises, it would be nevertheless useful to have more unified code packages where the majority of the most commonly used methods are implemented, as well as electronic structure codes where all possible quantities, such as NACVs\cite{Truhlar2023nacv}, can be calculated.

The MASH algorithm used in this work is currently only applicable to two-state systems and there is a need to generalize for multi-state problems. Two distinct approaches have emerged for generalizing MASH\cite{Runeson2023MASH,Runeson2024MASH,MSMASH}, both of which were applied to study the photochemistry of cyclobutanone in the recent JCP community challenge.\cite{cyclobutanone,hutton2024using} We wait to see how both ideas develop in the future. 

\rev{
\textbf{Supporting Information:}
The supplementary material provides all of the necessary information to reproduce the surface hopping and AIMS results in the main text. This includes the MOLPRO input and SHARC template files, the ground-state geometries and frequencies and a python script for computing the ethylene product yields.
}

\begin{acknowledgement}
This work was supported by the Cluster of Excellence ``CUI: Advanced Imaging of Matter'' of the Deutsche Forschungsgemeinschaft (DFG) – EXC 2056 – project ID 390715994. We would also like to thank Basile Curchod and Lea Ibele for useful discussions.
\end{acknowledgement}

\bibliography{references,extra_refs}

\end{document}


\maketitle

\renewcommand{\thepage}{S\arabic{page}}
\renewcommand{\theequation}{S\arabic{equation}}
\renewcommand{\thefigure}{S\arabic{figure}}
\renewcommand{\thetable}{S\arabic{table}}
\renewcommand{\thesection}{S\arabic{section}}
\renewcommand{\thesubsection}{S\arabic{section}.\arabic{subsection}}

Here we provide all of the necessary information to reproduce the results in the main text, including details on the electronic structure and dynamics methods employed, choosing the initial conditions for the simulations, and constructing the nuclear observables for the product yields.

\section{Electronic structure}
For ethylene and fulvene, the exited states are computed at the SA-CASSCF level of theory performed in MOLPRO 2012\cite{MOLPRO2012} with (2,2) and (6,6) active spaces respectively, which contain all of the $\pi$ and $\pi^{*}$ orbitals within the molecules. A 6-31G* basis set with spherical harmonic basis functions is used. For ethylene, the state-averaging procedure is performed over the three lowest energy singlet states, but the dynamics is only considered in the subspace of the lowest two states. While previous work on the three-state case has suggested that $\mathrm{S}_{2}$ can play an important role in the initial dynamics, we choose to stick with the simplicity of the two-state description for benchmarking purposes.\cite{Gomez2024} 

For DMABN, the $\mathrm{S}_{1}$ and $\mathrm{S}_{2}$ surfaces are computed using LR-TDDFT within the Tamm-Dancoff approximation using the LC-PBE functional and the 6-31G basis set. The TDDFT calculations were performed in GAUSSIAN 16\cite{g16} and the range-separation parameter for the functional was set to 0.3 Bohr. 

The SHARC template and AIMS/MOLPRO input files for each molecule, which uniquely specifies the electronic structure theory used, are also provided.

\subsection{Nonadiabatic Coupling Vectors}
For LR-TDDFT calculations in GAUSSIAN 16,\cite{g16} analytic NACVs are currently unavailable. The electronic propagation is therefore performed in a local diabatic basis,\cite{Plasser2016} which is also the default option within the SHARC\cite{SHARC2.0,Mai2018SHARC} package. A `crude adiabatic basis' is chosen, which is the adiabatic basis associated with the nuclear geometry at the beginning of each time step.  For the nuclear time step between $t$ and $t+\Delta t$, the electronic Hamiltonian at time $t$ is diagonal in this basis, while the Hamiltonian at time $t+\Delta t$ can be expressed as a similarity transform of the associated potential energies,
\begin{subequations}
\begin{align}
H_{\beta\gamma}(\bm{q}(t))&=V_{\beta}(\bm{q}(t))\delta_{\beta\gamma}, \label{eq:ham_init}  \\	
H_{\beta\gamma}(\bm{q}(t+\Delta t))&=\sum_{\alpha}S_{\beta\alpha}V_{\alpha}(\bm{q}(t+\Delta t))S^{\dagger}_{\alpha\gamma} , \label{eq:ham_final} \\
S_{\beta\alpha}&=\braket{\psi_{\beta}(\bm{q}(t))|\psi_{\alpha}(\bm{q}(t+\Delta t))}\mathrm{sgn}(\braket{\psi_{\alpha}(\bm{q}(t))|\psi_{\alpha}(\bm{q}(t+\Delta t))}) . \label{eq:overlap}
\end{align}
\end{subequations} 
The sign function ensures that the wavefunctions at the two times have a consistent sign, with the overall sign of the overlap matrix given by the sign of the wavefunction at time $t$. The sign of the wavefunction at time $t+\Delta t$ is then updated to be consistent with the sign of the wavefunction at time $t$, in order to calculate the overlap matrix at the next time step.

For the electronic propagation across the nuclear time step, $\Delta t$, the propagation is further split up into $n_{\mathrm{elec}}$ electronic time steps, $\delta t=\Delta t/n_{\mathrm{elec}}$ and the electronic Hamiltonian is linearly interpolated between Eq.~\ref{eq:ham_init} and Eq.~\ref{eq:ham_final}. The propagation of the electronic wavefunction coefficients is then given by
\begin{subequations}
\begin{align}
\tilde{H}_{\alpha\beta}(t_{k})&=H_{\alpha\beta}(\bm{q}(t))+\frac{k}{n_{\mathrm{elec}}}\left[H_{\alpha\beta}(\bm{q}(t+\Delta t))-H_{\alpha\beta}(\bm{q}(t))\right] \\
\tilde{c}_{\alpha}(t+\Delta t)&=\sum_{\beta}\left[\prod_{k=1}^{n_{\mathrm{elec}}}\eu{-\tfrac{i}{\hbar}\tilde{\bm{H}}(t_{k})\delta t}\right]_{\alpha\beta}\tilde{c}_{\beta}(t) , \label{eq:schrodinger}
\end{align}
\end{subequations}
where $\eu{-i\tilde{\bm{H}}(t_{k})}$ is a matrix exponential, $\tilde{c}_{\alpha}$ are the electronic wavefunction coefficients in the crude adiabatic basis and $\tilde{c}_{\alpha}(t)=c_{\alpha}(t)$. In this work, we use $n_{\mathrm{elec}}=25$. Once the electronic wavefunction has been successfully propagated to $t+\Delta t$, it is then transformed back into the adiabatic basis as follows 
\begin{equation}
c_{\alpha}(t+\Delta t)=\sum_{\beta}S_{\alpha\beta}^{\dagger}\,\tilde{c}_{\beta}(t+\Delta t) .
\end{equation}
Given that the couplings in the diabatized Hamiltonian are generally less localized compared to the NACVs, we use this electronic propagation scheme even when analytic NACVs are available.

In order to rescale the velocity when analytic NACVs are not available, we use the following finite difference scheme
\begin{eqnarray}\nonumber
d_{j}(\bm{q})&\approx&\frac{1}{2\Delta x} \Big( \braket{\psi_{+}(\bm{q})|\psi_{-}(\bm{q}+\Delta x\bm{j})}\mathrm{sgn}(\braket{\psi_{-}(\bm{q})|\psi_{-}(\bm{q}+\Delta x\bm{j})}) \\ &-& \braket{\psi_{+}(\bm{q})|\psi_{-}(\bm{q}-\Delta x\bm{j})}\mathrm{sgn}(\braket{\psi_{-}(\bm{q})|\psi_{-}(\bm{q}-\Delta x\bm{j})}) \Big) \label{eq:NACV}
\end{eqnarray}
where $\bm{j}$ is a unit vector pointing along the Cartesian coordinate associated with nuclear degree of freedom, $j$. Because the NACV is only used to determine the direction of the velocity rescaling, the signs of the NACVs at different time steps do not have to be consistent. The NACV is only computed at time steps where either a hop or frustrated hop have taken place and we use $\Delta x=0.001$~Bohr in all of our simulations.

In order to test this finite difference scheme, $\sum_{j}d_{j}(\bm{q}(t))v_{j}(t)$ can be calculated in the small nuclear time step limit as
\begin{equation}
\sum_{j}d_{j}(\bm{q}(t))v_{j}(t)\approx\frac{1}{\Delta t}\braket{\psi_{+}(\bm{q}(t))|\psi_{-}(\bm{q}(t+\Delta t))}\mathrm{sgn}(\braket{\psi_{-}(\bm{q}(t))|\psi_{-}(\bm{q}(t+\Delta t))}) , \label{eq:d.p}
\end{equation}
and compared with the same quantity computed using Eq.~\ref{eq:NACV}. Additionally, Eq.~\ref{eq:d.p} was used to calculate $\sum_{j}d_{j}(\bm{q}(t))v_{j}(t)$  for DMABN in Fig. 1 of the main paper.

For both the finite difference scheme for the NACVs and the local diabatic electronic propagation scheme, the required wavefunction overlaps were computed using the associated code in SHARC.\cite{Plasser2016} In addition, a L\"owdin orthogonalization is performed on all overlap matrices to ensure unitarity. In the case of ethylene, this orthogonalization procedure is performed in the three-state space.

\section{Dynamics Methods}
\subsection{Ab Initio Multiple Spawning (AIMS)}
Our AIMS simulations for ethylene and fulvene were performed using the AIMS/MOLPRO code.\cite{AIMS/MOLPRO} A time step of 20 a.u. was used outside the nonadiabatic coupling region. The coupling region is entered when $\hbar\sum_{j}d_{j}v_{j}>0.005\,E_{h}$, for which the time step was reduced to 5 a.u.. The value of this coupling threshold parameter was determined by reducing its value until there was no noticeable change in the obtained results. 

In the coupling region, Gaussians can be spawned whenever $\sum_{j}d_{j}v_{j}$ reaches a maximum along the trajectory. For this to occur, the overlap between the parent and child Gaussians at the spawning point must also exceed a threshold value, set for our calculations to 0.6. A minimum population of 0.01 is also required for a Gaussian to spawn a child.

The threshold for energy violation over a time step is set for ethylene and fulvene as 0.03 and 0.01 $E_{h}$ respectively. If this is violated, the time step is halved unless the minimum value of 1 a.u. is reached, after which energy violation is ignored. Ad hoc features of the AIMS simulations, such as setting a decoherence time, are not used in our simulations in order to obtain the most accurate and rigorous AIMS benchmark for these systems. 

When performing AIMS simulations, it is important that the initial nuclear geometry in the MOLPRO input and Geometry.dat files exactly match for each trajectory. The initial momenta in the Geometry.dat file must be provided in atomic units. This can be obtained from the initial velocities used in SHARC by multiplying them by the appropriate atomic mass in atomic units.

For the AIMS simulations, 400 and 150 trajectories were run for ethylene and fulvene respectively. Of these, 8 trajectories failed for ethylene due to a failure in converging the CASSCF cycle. We did not perform the AIMS calculation for DMABN ourselves and the results were instead taken from Ref.~\onlinecite{Curchod2017}. The details for the AIMS simulation of DMABN can therefore be found there.  

\begin{figure}[h]
    \centering
    \includegraphics[width=\linewidth]{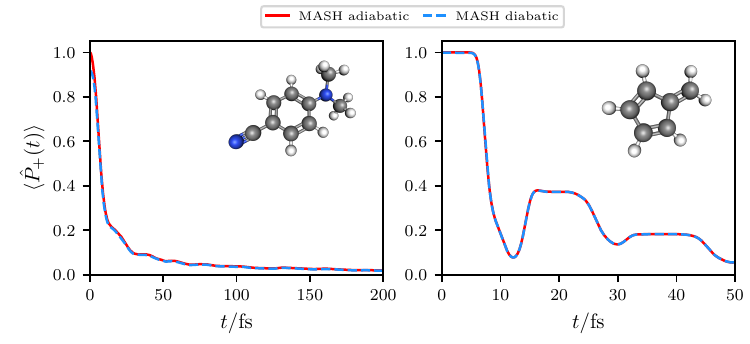}
    \caption{Figure showing the MASH dynamical populations of the upper adiabatic state in the LVC models, when starting in either an adiabatic or diabatic state. The MASH results when starting in an initial diabatic state are the same as given in Fig.~3 of the main paper.}
    \label{fig:initial}
\end{figure}

For the linear vibronic coupling (LVC) models, the AIMS dynamics were performed with a modified version of the FMS90 code implemented in MOLPRO.\cite{AIMS_CI} 2000 AIMS trajectories were run with a times step of 2.0 a.u., reduced to 0.5 a.u. in the coupling region. The coupling region was entered when $\hbar\sum_{j}d_{j}v_{j}>0.002\,E_{h}$ and a minimum population of 0.001 and an overlap threshold of 0.6 were set for spawning Gaussians. All other parameters were kept at their default value. For AIMS simulations, the system has to be initialized in an adiabatic state, whereas for all other approaches in the LVC models, we initialized the dynamics in a diabatic state. Fig.~\ref{fig:initial} gives the MASH adiabatic populations associated with starting in an adiabatic and diabatic state, which illustrates that apart from the $t=0$ value in the DMABN model, the initial conditions do not make much of a difference to the obtained populations in these LVC models.

As is standard practice in AIMS, the electronic populations are calculated using the overlaps between the Gaussians\cite{Curchod2017} 
\begin{equation}
\braket{\hat{P}_{+}(t)}=\frac{1}{N_{\mathrm{runs}}}\sum_{n=1}^{N_{\mathrm{runs}}}\sum_{k,l=1}^{N_{\mathrm{Gauss},n}(t)}C^{*}_{k,n,+}(t)C_{l,n,+}(t)\braket{g_{k,n}(\bm{q}_{k,n}(t),\bm{p}_{k,n}(t))|g(\bm{q}_{l,n}(t),\bm{p}_{l,n}(t))} ,
\end{equation}
where $N_{\mathrm{runs}}$ is the number of independent AIMS runs and $N_{\mathrm{Gauss,n}}(t)$ is the number of Gaussians in run $n$ at time $t$. Additionally, $q_{k,n}(t)$ and $p_{k,n}(t)$ are the classical nuclear phase-space variables at which the Gaussian $g_{k,n}$ is centered and $C_{k,n,+}(t)$ is its time-dependent weight, where the $+$ subscript indicates that only Gaussians on the upper surface are used to compute the population of the upper adiabat.

In contrast for nuclear observables like the DMABN twist angle and the ethylene product yields, only the diagonal elements of the Gaussian basis are used. For observables $\hat{O}$ that are functions solely of the nuclear positions, $\hat{\bm{q}}$\cite{Curchod2017}
\begin{equation}
\braket{\hat{O}(t)}\approx\frac{1}{N_{\mathrm{runs}}}\sum_{n=1}^{N_{\mathrm{runs}}}\frac{\sum_{k=1}^{N_{\mathrm{Gauss},n}}|C_{k,n}(t)|^{2}\,O(\bm{q}_{k,n}(t))}{\sum_{k=1}^{N_{\mathrm{Gauss},n}}|C_{k,n}(t)|^{2}} ,
\end{equation}
where the lack of the $+$ subscript on the Gaussian weights now indicates that all Gaussians contribute to the observable, irrespective of the surface they are on. 
\subsection{Fewest-Switches Surface Hopping (FSSH)}
In FSSH,\cite{Subotnik2016review} the nuclei are propagated according to Newton's equations of motion on the active Born-Oppenheimer surface and the electronic wavefunction is propagated according to the time-dependent Schr\"odinger equation [Eq.~\ref{eq:schrodinger}]. The velocity-verlet scheme is used for the nuclear dynamics with $\Delta t = 0.25$ fs. In order to initialize the electronic subsystem in the excited state, the initial wavefunction coefficients are given by $c_{-}(0)=0$ and $c_{+}(0)=1$.

In order to describe nonadiabatic transitions, the active surface is changed stochastically with the following probability
\begin{equation}	P_{\alpha\rightarrow\beta}=\frac{2\mathrm{Re}\left[c^{*}_{\alpha}S_{\alpha\beta}c_{\beta}\right]}{|c_{\alpha}|^{2}} ,
\end{equation}
where $\mathrm{Re}[\cdot]$ returns the real part of the quantity and any negative probabilities are set to zero. Electronic population observables are constructed in terms of the active state, rather than the electronic wavefunction coefficients.

At an attempted hop, the kinetic energy along the NACV must first be compared with the electronic transition energy.  A hop from state $\alpha$ to $\beta$ is energetically allowed if
\begin{equation}
\label{eq:hop_cond}
\frac{\left(\sum_{j}d_{j}v_{j}\right)^{2}}{2\sum_{j}\frac{d_{j}^{2}}{m_{j}}}>V_{\beta}-V_{\alpha} , 
\end{equation}
in which case the velocity along the NACV is rescaled according to
\begin{equation}
v_{n}^{\mathrm{new}}=v_{n}^{\mathrm{old}}-\frac{\frac{d_{n}}{m_{n}}}{\sum_{j}\frac{d_{j}^{2}}{m_{j}}}\left[\sum_{j}d_{j}v_{j}^{\mathrm{old}}-\mathrm{sgn}\left(\sum_{j}d_{j}v_{j}^{\mathrm{old}}\right)\sqrt{\left(\sum_{j}d_{j}v_{j}^{\mathrm{old}}\right)^{2}+2(V_{\alpha}-V_{\beta})\left(\sum_{j}\frac{d_{j}^{2}}{m_{j}}\right)}\right] .
\end{equation}
In the event that Eq.~\ref{eq:hop_cond} is not satisfied, the hop is frustrated and the nuclear velocity along the NACV is reflected as follows
\begin{equation}
	v_{n}^{\mathrm{new}}=v_{n}^{\mathrm{old}}-2\left(\sum_{j}d_{j}v_{j}^{\mathrm{old}}\right)\frac{\frac{d_{n}}{m_{n}}}{\sum_{j}\frac{d_{j}^{2}}{m_{j}}} .
\end{equation}
This treatment of frustrated hops in FSSH required a modification to the SHARC code in order to ensure that the nuclear velocity is always reflected at a frustrated hop. 

To add decoherence into FSSH simulations, we use the energy-based decoherence scheme described in Ref.~\onlinecite{Granucci2007,Granucci2010}. For a trajectory with current active surface $\beta$, the electronic coefficients are updated at each time step as follows
\begin{subequations}
\begin{align}
c^{\mathrm{new}}_{\alpha}=&c_{\alpha}^{\mathrm{old}}\eu{-\tfrac{1}{2}\Delta t \frac{|V_{\alpha}-V_{\beta}|}{\hbar\left(1 + \frac{C}{E_{\mathrm{kin}}}\right)}} ,  \\
c^{\mathrm{new}}_{\beta}=&\frac{c_{\beta}^{\mathrm{old}}}{|c_{\beta}^{\mathrm{old}}|}\sqrt{1 - \sum_{\alpha\neq\beta}|c_{\alpha}^{\mathrm{new}}|^{2}} ,
\end{align}
\end{subequations}
where $E_{\mathrm{kin}}$ is the total nuclear kinetic energy and  $C=0.1\,E_{h}$ is the decoherence parameter. 

All FSSH simulations were performed using SHARC 2.0. \cite{Mai2018SHARC} For ethylene, DMABN and fulvene, 1000, 600 and 600 trajectories were run respectively to ensure well converged results. However far fewer trajectories would be needed for a good qualitative description of the observables. For ethylene only 5 trajectories or fewer failed as a result of being unable to converge the CASSCF cycle, whereas for fulvene it was 3 trajectories or fewer.

For the LVC models, the surface hopping approaches were initialized in a diabatic state, to allow direct comparison with the MCTDH and vMCG results. Ref.~\onlinecite{MASH} explains how to compute diabatic populations with FSSH. We choose to use a very large number of trajectories (100000) to essentially remove any statistical error within the results, which is easy to do in model calculations. However, far fewer trajectories could have been used in practice to qualitatively reproduce the correct dynamical behaviour.

\subsection{A Mapping Approach to Surface Hopping (MASH)}
In order to perform \textit{ab initio} MASH simulations, a modified version of SHARC was used. Due to the similarities between the FSSH and MASH algorithms, only minor modifications to the existing surface-hopping routine were required, which we describe here.

The first difference between FSSH and MASH is the initial sampling of the electronic wavefunction. When starting in the upper adiabatic state, the electronic wavefunction in MASH is sampled from the upper hemisphere of the Bloch sphere.\cite{Mannouch2023MASH} This corresponds to
\begin{subequations}
\begin{align}
c_{+}&=\cos(\frac{\theta}{2})\eu{-i\phi/2} , \\
c_{-}&=\sin(\frac{\theta}{2})\eu{i\phi/2} ,
\end{align}
\end{subequations}
with $0\leq\theta<\pi/2$ and $0\leq\phi< 2\pi$. The sampling over the upper hemisphere of the Bloch sphere is not uniform however, but is weighted by the factor: $|c_{+}|^{2}-|c_{-}|^{2}$. To incorporate this, the values of $\theta$ and $\phi$ are sampled according to
\begin{subequations}
\begin{align}
\theta=&\cos^{-1}(\sqrt{a}) , \\
\phi=&2\pi b , 
\end{align}
\end{subequations}
where $a$ and $b$ are themselves sampled uniformally and independently between the values 0 and 1. In SHARC, this can be easily implemented by sampling the initial electronic wavefunction externally and then reading it in through an external coefficient file.

The second difference is that in MASH, the active surface is no longer a stochastic variable, but is uniquely determined from the electronic wavefunction.\cite{Mannouch2023MASH} The active surface variable, $n_{\mathrm{active}}$, in MASH is given by
\begin{equation}
n_{\mathrm{active}}=\mathrm{sgn}(|c_{+}|^{2}-|c_{-}|^{2}) ,
\end{equation}
where $n_{\mathrm{active}}=1$ corresponds to propagation on the upper surface and $n_{\mathrm{active}}=-1$ on the lower surface.

All MASH simulations were performed using a locally modified (as described above) version of SHARC 2.0\cite{Mai2018SHARC}. For ethylene, DMABN and fulvene, 1000, 600 and 600 trajectories were run respectively to ensure well converged results. However far fewer trajectories would be needed for a good qualitative description of the observables. For ethylene, only 4 trajectories failed as a result of being unable to converge the CASSCF cycle.

For the LVC models, MASH was initialized in a diabatic state, to allow direct comparison with the MCTDH and vMCG results. Ref.~\onlinecite{MASH} explains how to do this with MASH, as well as how to compute diabatic populations. We choose to use a very large number of trajectories (100000) to essentially remove any statistical error within the results, which is easy to do in model calculations. However, far fewer trajectories could have been used in practice to qualitatively reproduce the correct dynamical behaviour.

\section{Nuclear Initial Conditions Sampling}
The nuclei were initialized in the nuclear ground-state associated with the ground-state Born-Oppenheimer surface. As is common practice, this was approximated as a multidimensional harmonic oscillator. The optimization of the ground-state geometry, $q_{j}'$ and frequencies, $\omega_{j}$, were performed using the same level of electronic structure as the dynamics, with the atomic masses chosen to be consistent with the AIMS/MOLPRO code.\cite{AIMS/MOLPRO} These are given in Table.~\ref{tab:masses} in both atomic units (a.u.) and atomic mass units (a.m.u.). For each system, the MOLDEN files containing the equilibrium ground-state geometry and frequencies are also provided. 
\begin{table*}
	\caption{Atomic masses used for the \textit{ab initio} simulations performed in this work, given in atomic units (a.u.) and atomic mass units (a.m.u.).
	}\label{tab:masses}
	\begin{tabular}{lccc}
		Atom & \ch{C} & \ch{H} & \ch{N}  \\ 
		\hline
		Mass (a.u.)  & 21874.644 & 1822.887 & 25520.418   \\ 
		Mass (a.m.u.)  & 11.999990224335 & 0.999999185361 & 13.999988595055 
	\end{tabular}
\end{table*}

For the dynamical approaches considered in this work, the initial nuclear phase-space variables for the trajectories are sampled from the Wigner distribution of the ground-state quantum harmonic oscillator
\begin{equation}
	\label{eq:wigner_dist}
	\rho_{\mathrm{nuc}}(\tilde{\bm{q}},\tilde{\bm{p}})=\prod_{j=1}^{N}\,\frac{\hbar}{\pi}\,\mathrm{exp}\left[-\frac{\tilde{p}_{j}^{2}+\omega_{j}^{2}\left(\tilde{q}_{j}-\tilde{q}'_{j}\right)^{2}}{\hbar\omega_{j}}\right] ,
\end{equation}
where the tildes signify mass-weighted coordinates. In SHARC, this sampling can be performed using the command
\begin{verbatim}
$SHARC/wigner.py -n 600 -m freq.molden
\end{verbatim}
where the -m option allows the use of non-standard masses, the -n option specifies the number of samples (in this case 600) and freq.molden is the MOLDEN file containing the frequencies and the ground-state geometry.

Not all previous simulations on these systems used the same nuclear initial conditions however.  For the AIMS simulation of DMABN in Ref.~\onlinecite{Curchod2017}, initial Wigner sampled phase-space variables were further refined according to their associated oscillator strength for the  $\mathrm{S}_{0}\rightarrow\mathrm{S}_{2}$ photoexcitation transition. For the surface-hopping algorithms, we tested the initial nuclear conditions with and without this further refinement and found that it only made negligible differences to the results. We therefore chose to present the results corresponding to sampling from the Wigner distribution given by Eq.~\ref{eq:wigner_dist}.

Additionally in Ref.~\onlinecite{Ibele2020}, the dynamical simulations of fulvene set the initial nuclear velocity to zero, in order to ensure that trajectories only initially passed through the sloped conical intersection seam. We however find that this is also the case when trajectories are sampled from the initial Wigner distribution, as well as having the additional advantage of corresponding to a valid quantum state. 

\section{Nuclear Observables}
\subsection{Product Yields in Ethylene}
In order to define the various products arising from the photoexcitation of ethylene, certain bond lengths must first be defined. Most straightforwardly, the carbon-carbon bond length, $r_{\mathrm{C}-\mathrm{C}}$, is defined as the distance between the two carbon atoms, $\mathrm{C}_{1}$ and $\mathrm{C}_{2}$. For each hydrogen atom, $\mathrm{H}_{i}$, the corresponding carbon-hydrogen bond length, $r_{\mathrm{C}-\mathrm{H}_{i}}$, is taken as the smallest of $r_{\mathrm{C}_{1}-\mathrm{H}_{i}}$ and $r_{\mathrm{C}_{2}-\mathrm{H}_{i}}$. Finally, we define $\mathrm{H}_{4}$ and $\mathrm{H}_{3}$ as the two hydrogen atoms with the largest and second largest carbon-hydrogen bond lengths respectively (and are therefore the hydrogens that have either dissociated or are closest to doing so). The hydrogen-hydrogen bond distance, $r_{\mathrm{H}-\mathrm{H}}$, is then defined as the distance between $\mathrm{H}_{4}$ and $\mathrm{H}_{3}$.  

\begin{figure}[t]
    \centering
    \includegraphics[width=0.8\linewidth]{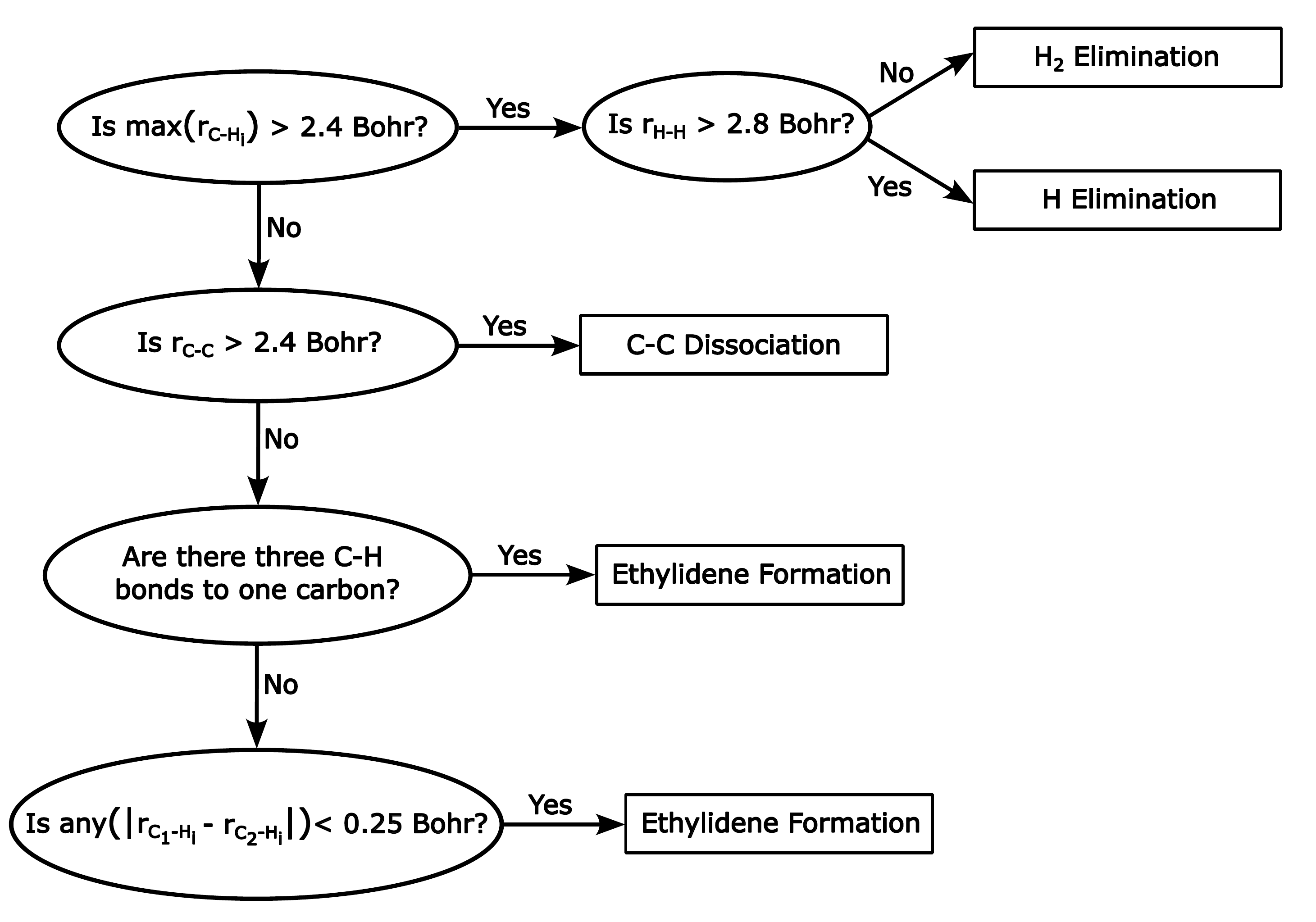}
    \caption{A flowchart defining the various products formed after the photoexcitation of ethylene. The first row of the diagram determines whether any C-H bonds have been cleaved, the second whether the C-C bond has been broken, the third whether any hydrogen atoms have migrated to another carbon and the fourth looks for any hydrogen bridging atoms.}
    \label{fig:yield_define}
\end{figure}
Figure.~\ref{fig:yield_define} contains a flowchart that shows how we determined the products formed at a given time step of a trajectory. This was based on the scheme used in Ref.~\onlinecite{Barbatti2005} and was tested by analyzing a sufficiently large random selection of trajectories by hand. The bond-length criteria used in our product specifications are very different from the equilibrium bond lengths in ethylene, because ethylene is highly vibrationally excited once it relaxes back to the electronic ground state after the initial electronic photoexcitation.  

\subsection{The Twist Angle of the Dimethylamino Group in DMBAN}
\begin{figure}[t]
    \centering
    \includegraphics[width=0.7\linewidth]{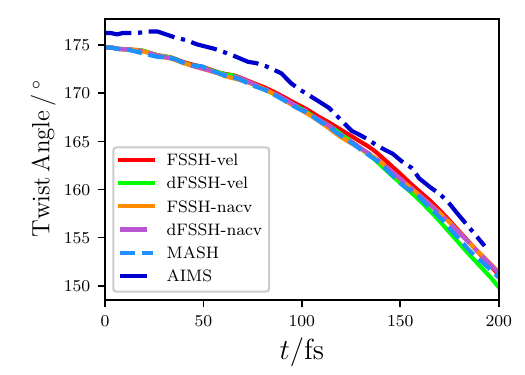}
    \caption{The twist angle of the dimethylamino group in DMABN. The AIMS result is taken from Ref.~\protect\onlinecite{Curchod2017}.}
    \label{fig:twist}
\end{figure}
We use the same definition of the twist angle of the dimethylamino group in DMABN as Eq.~7 in Ref.~\onlinecite{Curchod2017}.
Figure~\ref{fig:twist} gives the twist angle of the dimethylamino group in DMABN as a function of time. The AIMS result was taken from  Ref.~\protect\onlinecite{Curchod2017}. Because only 20 AIMS trajectories were used to calculate the twist angle (compared to 200 trajectories for the surface hopping results), we expect that the small difference between the AIMS and surface hopping twist angles is due to statistical error in the former. 

\section{Additional results}
\begin{figure}[h]
    \centering
    \includegraphics[width=0.5\linewidth]{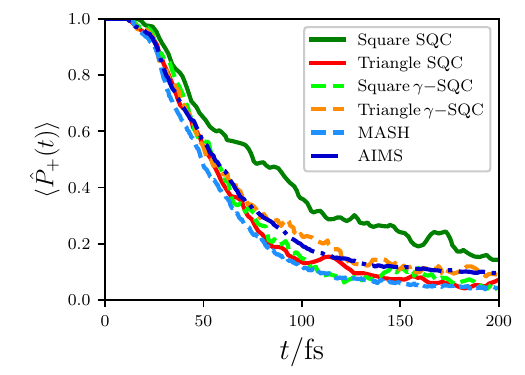}
    \caption{Figure showing the dynamical population of the upper adiabatic state in ethylene. The SQC results were taken from Ref.~\protect\onlinecite{Weight2021}.}
    \label{fig:Ethylene_SQC}
\end{figure}

In Fig.~\ref{fig:Ethylene_SQC} we compare SQC results from Ref.\onlinecite{Weight2021} with our AIMS and MASH results for the excited state population of ethylene.

\begin{figure}[h]
    \centering
    \includegraphics[width=0.8\linewidth]{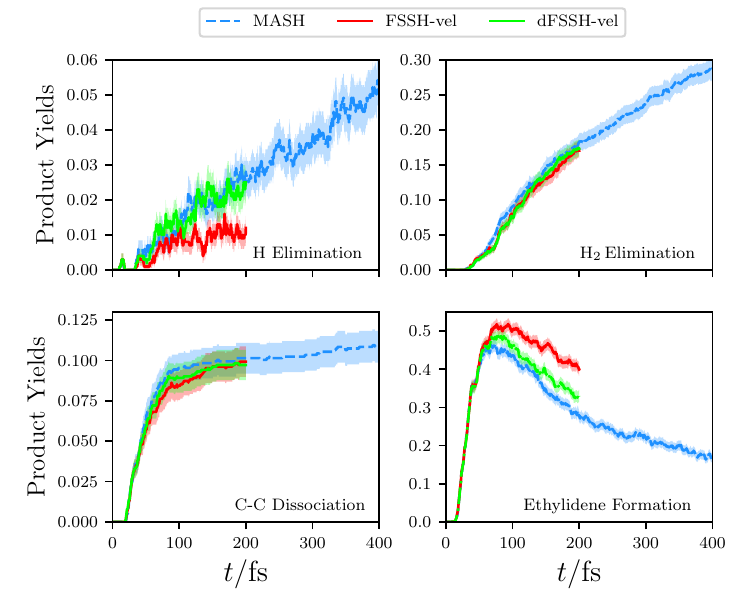}
    \caption{Dynamical product yields for the major products in the photodissociation of ethylene, calculated using FSSH-vel and dFSSH-vel. The MASH result is the same as in Fig.~4 of the main paper. The width of the shading represents twice the standard error.}
    \label{fig:yield_supp}
\end{figure}

Additionally, Fig.~\ref{fig:yield_supp} gives the dynamical product yields calculated for FSSH-vel and dFSSH-vel, which were not included in the main paper. Interestingly unlike FSSH-nacv, FSSH-vel is able to accurately reproduce the product yield for $\mathrm{H}_{2}$ elimination, but instead under predicts the product yield for H elimination.

\section{Nuclear Momenta in the Adiabatic Basis}
In this section, we define the kinematic and canonical nuclear momenta in the adiabatic basis for quantum wavepacket approaches. An equivalent definition more applicable to semiclassical trajectory approaches can be found in Ref.~\onlinecite{Cotton2017mapping}.

In order to keep the resulting equations as simple as possible, we start with the general form of the Hamiltonian for an electron-nuclear coupled system expressed in a diabatic electronic basis
\begin{subequations}
\begin{align}
\hat{H}&=\sum_{j}\frac{\hat{p}_{j}^{2}}{2m_{j}}+\hat{V}(\hat{\bm{q}}) , \\
\hat{V}(\hat{\bm{q}})&=\sum_{kl}V_{kl}(\hat{\bm{q}})\ket{k}\bra{l} .
\end{align}
\end{subequations}
Here, $\hat{p}_{j}$ is the momentum operator for nuclear degree of freedom $j$ of mass $m_{j}$ and $\hat{V}(\hat{\bm{q}})$ is the potential matrix evaluated in a diabatic basis, $\ket{k}$, which is independent of the nuclear coordinates. From this expression for the Hamiltonian, the equation of motion for the expectation value of the nuclear momentum operator can be obtained as follows
\begin{equation}
\label{eq:moment_eom}
\begin{split}
\frac{\rd}{\rd t}\braket{\hat{p}_{j}}&=\frac{i}{\hbar}\braket{[\hat{H},\hat{p}_{j}]} \\
&=-\Braket{\hat{V}'(\hat{\bm{q}})} ,
\end{split}
\end{equation}
where 
\begin{equation}
\hat{V}'(\hat{\bm{q}})=\sum_{kl}V'_{kl}(\hat{\bm{q}})\ket{k}\bra{l} ,
\end{equation}
and $V'_{kl}(\bm{q})=\frac{\partial V_{kl}(\bm{q})}{\partial q_{j}}$.

In order to obtain an analogous expression for the adiabatic basis, we express the full quantum state of the coupled electron-nuclear system using the Born-Huang expansion
\begin{equation}
\ket{\psi(t)}=\sum_{\lambda}\int\rd\bm{q}\,\chi^{(\lambda)}(\bm{q},t)\ket{\psi_{\lambda}(\bm{q})}\ket{\bm{q}} ,
\end{equation}
where $\ket{\psi_{\lambda}(\bm{q})}$ is the electronic state associated with adiabat $\lambda$ at nuclear coordinate $\bm{q}$, $\ket{\bm{q}}$ is the eigenstate of the nuclear position operator and $\chi^{(\lambda)}(\bm{q},t)$ is the nuclear wavefunction associated with adiabat $\lambda$. For example in AIMS, $\chi^{(\lambda)}(\bm{q},t)$ takes the form of a linear combination of Gaussian functions.

In order to evaluate the right-hand side of Eq.~(\ref{eq:moment_eom}) in the adiabatic basis, for which $\hat{V}(\bm{q})\ket{\psi_{\lambda}(\bm{q})}=V_{\lambda}(\bm{q})\ket{\psi_{\lambda}(\bm{q})}$, we can use the fact that
\begin{equation}
\braket{\psi_{\mu}(\bm{q})|\hat{V}'(\bm{q})|\psi_{\lambda}(\bm{q})}=\frac{\partial V_{\lambda}(\bm{q})}{\partial q_{j}}\delta_{\mu\lambda}-V_{\lambda}(\bm{q})\Braket{\frac{\partial \psi_{\mu}(\bm{q})}{\partial q_{j}}|\psi_{\lambda}(\bm{q})}-V_{\mu}(\bm{q})\Braket{\psi_{\mu}(\bm{q})|\frac{\partial \psi_{\lambda}(\bm{q})}{\partial q_{j}}} .
\end{equation}
Additionally using $d_{j}^{(\mu,\lambda)}(\bm{q})=\Braket{\psi_{\mu}(\bm{q})|\frac{\partial \psi_{\lambda}(\bm{q})}{\partial q_{j}}}=-\Braket{\frac{\partial\psi_{\mu}(\bm{q})}{\partial q_{j}}|\psi_{\lambda}(\bm{q})}$ leads to the expression given by Eq.~(1) in the main paper.

Finally, the left-hand side of Eq.~\ref{eq:moment_eom} can be evaluated in the adiabatic basis to give 
\begin{equation}
\label{eq:kinematic_def}
\braket{\hat{p}_{j}}=\sum_{\lambda}\braket{\chi^{(\lambda)}(t)|\hat{p}_{j}|\chi^{(\lambda)}(t)}+2\hbar\sum_{\lambda}\sum_{\mu>\lambda}\mathrm{Im}\left[\braket{\chi^{(\mu)}(t)|d_{j}^{(\mu,\lambda)}(\hat{\bm{q}})|\chi^{(\lambda)}(t)}\right]
\end{equation}
where $\ket{\chi^{(\lambda)}(t)}=\int\rd\bm{q}\,\chi^{(\lambda)}(\bm{q},t)\ket{\bm{q}}$. The first term on the right-hand side of this expression corresponds to the nuclear momenta associated with the nuclear wavefunctions on each adiabat, which is referred to as the conjugate momenta in the adiabatic basis. The total momenta, which also contains a contribution arising from the nuclear coordinate dependence of the adiabatic states, is referred to as the kinematic momenta. Defining $\braket{\hat{p}^{\mathrm{ad}}_{j}}=\sum_{\lambda}\braket{\chi^{(\lambda)}(t)|\hat{p}_{j}|\chi^{(\lambda)}(t)}$, Eq.~(\ref{eq:kinematic_def}) can also be expressed in operator form as 
\begin{equation}
\hat{p}_{j}=\hat{p}_{j}^{\mathrm{ad}}+\hbar\sum_{\lambda}\sum_{\mu>\lambda}d_{j}^{(\mu,\lambda)}(\hat{\bm{q}})\hat{\sigma}_{y}^{(\mu,\lambda)}(\hat{\bm{q}}) ,   
\end{equation}
where $\hat{\sigma}^{(\mu,\lambda)}_{y}(\hat{\bm{q}})=-i(\ket{\psi_{\mu}(\hat{\bm{q}})}\bra{\psi_{\lambda}(\hat{\bm{q}})}-\ket{\psi_{\lambda}(\hat{\bm{q}})}\bra{\psi_{\mu}(\hat{\bm{q}})})$. This means that we could have equivalently derived Eq.~(1) in the main paper using the adiabatic basis from the start as follows
\begin{subequations}
\begin{align}
&\frac{\rd}{\rd t}\Braket{\hat{p}^{\mathrm{ad}}_{j}+\hbar\sum_{\lambda}\sum_{\mu>\lambda}d_{j}^{(\mu,\lambda)}(\hat{\bm{q}})\hat{\sigma}_{y}^{(\mu,\lambda)}(\hat{\bm{q}})}=\frac{i}{\hbar}\Braket{\left[\hat{H}^{\mathrm{ad}},\hat{p}^{\mathrm{ad}}_{j}+\hbar\sum_{\lambda}\sum_{\mu>\lambda}d_{j}^{(\mu,\lambda)}(\hat{\bm{q}})\hat{\sigma}_{y}^{(\mu,\lambda)}(\hat{\bm{q}})\right]} , \\
&\hat{H}^{\mathrm{ad}}=\sum_{j}\frac{\left(\hat{p}^{\mathrm{ad}}_{j}+\hbar\sum_{\lambda}\sum_{\mu>\lambda}d_{j}^{(\mu,\lambda)}(\hat{\bm{q}})\hat{\sigma}_{y}^{(\mu,\lambda)}(\hat{\bm{q}})\right)^{2}}{2m_{j}}+\sum_{\lambda}V_{\lambda}(\hat{\bm{q}})\hat{P}_{\lambda}(\hat{\bm{q}}) .
\end{align}
\end{subequations}

\newpage

\clearpage
\bibliographystyle{achemso}
\bibliography{references,extra_refs}